\documentclass[aps,prb,twocolumn,showpacs,showkeys,superscriptaddress]{revtex4-1}

\usepackage{graphicx}% Include figure files
\usepackage{dcolumn}% Align table columns on decimal point
\usepackage{bm}% bold math
\usepackage{multirow}
\usepackage{amsmath,amssymb}
\usepackage{graphicx}
\usepackage{epstopdf}
\usepackage{ulem}
\usepackage{color}
\usepackage{subfigure}
\usepackage{lineno}

\definecolor{red}{rgb}{1,0,0}
\definecolor{blue}{rgb}{0,0,1}
\definecolor{black}{rgb}{0,0,0}

\begin{document}

% Use the \preprint command to place your local institutional report
% number in the upper righthand corner of the title page in preprint mode.
% Multiple \preprint commands are allowed.
% Use the 'preprintnumbers' class option to override journal defaults
% to display numbers if necessary
%\preprint{}

%Title of paper
%\title{Superconductivity in the topological superconductor candidate La$_{3}$Pt$_{3}$Bi$_{4}$}
\title{Superconductivity in TlBi$_2$ with a large Kadowaki-Woods ratio}

% repeat the \author .. \affiliation  etc. as needed
% \email, \thanks, \homepage, \altaffiliation all apply to the current
% author. Explanatory text should go in the []'s, actual e-mail
% address or url should go in the {}'s for \email and \homepage.
% Please use the appropriate macro foreach each type of information

% \affiliation command applies to all authors since the last
% \affiliation command. The \affiliation command should follow the
% other information
% \affiliation can be followed by \email, \homepage, \thanks as well.

\author{Zhihua Yang}
\affiliation {Hangzhou Key Laboratory of Quantum Matter, School of Physics, Hangzhou Normal University, Hangzhou 311121, China}

\author{Zhen Yang}
\affiliation {Hangzhou Key Laboratory of Quantum Matter, School of Physics, Hangzhou Normal University, Hangzhou 311121, China}

\author{Qiping Su}
\affiliation {Hangzhou Key Laboratory of Quantum Matter, School of Physics, Hangzhou Normal University, Hangzhou 311121, China}

\author{Jianhua Du}
\affiliation {Department of Physics, China Jiliang University, Hangzhou 310018, China}

\author{Enda Fang}
\affiliation {Hangzhou Key Laboratory of Quantum Matter, School of Physics, Hangzhou Normal University, Hangzhou 311121, China}

\author{Chunxiang Wu}
\affiliation {Department of Physics, Zhejiang University, Hangzhou 310027, China}

\author{Jinhu Yang}
\affiliation {Hangzhou Key Laboratory of Quantum Matter, School of Physics, Hangzhou Normal University, Hangzhou 311121, China}

\author{Bin Chen}
\affiliation {Hangzhou Key Laboratory of Quantum Matter, School of Physics, Hangzhou Normal University, Hangzhou 311121, China}

\author{Hangdong Wang}
\email{hdwang@hznu.edu.cn}
\affiliation {Hangzhou Key Laboratory of Quantum Matter, School of Physics, Hangzhou Normal University, Hangzhou 311121, China}

\author{Minghu Fang}
\email{mhfang@zju.edu.cn}
\affiliation {Department of Physics, Zhejiang University, Hangzhou 310027, China}
\affiliation {Collaborative Innovation Center of Advanced Microstructures,  Nanjing University, Nanjing 210093, China}

%\email[]{Your e-mail address}
%\homepage[]{Your web page}
%\thanks{}
%\altaffiliation{}

%Collaboration name if desired (requires use of superscriptaddress
%option in \documentclass). \noaffiliation is required (may also be
%used with the \author command).
%\collaboration can be followed by \email, \homepage, \thanks as well.
%\collaboration{}
%\noaffiliation

\date{\today}
%\linenumbers
\begin{abstract}
In this article, the superconducting and normal state properties of TlBi$_2$ with the AlB$_2$-type structure were studied by the resistivity, magnetization and specific heat measurements. It was found that bulk superconductivity with $T_{C}$ = 6.2 K emerges in TlBi$_2$, which is a phonon-mediated $s$-wave superconductor with a strong electron-phonon coupling ($\lambda$$_{ep}$ = 1.38) and a large superconducting gap ($\Delta_{0}$/$k_{B}T_{C}$ = 2.25). We found that the $\rho$($T$) exhibits an unusual $T$-linear dependence above 50 K, and can be well described by the Fermi-liquid theory below 20 K. Interestingly, its Kadowaki-Woods ratio $A/\gamma^{2}$ [9.2$\times$10$^{-5}$ $\mu\Omega$ cm(mol K$^{2}$/mJ)$^{2}$] is unexpectedly one order of magnitude larger than that obtained in many heavy Fermi compounds, although the electron correlation is not so strong.

%Our results suggest that TlBi$_{2}$ compound is a good comparative study object of MgB$_{2}$ superconductor.

\end{abstract}

% insert suggested PACS numbers in braces on next line
% insert suggested keywords - APS authors don't need to do this

%\maketitle must follow title, authors, abstract, \pacs, and \keywords
\maketitle

\section{INTRODUCTION}
MgB$_{2}$, as a simple binary compound, has a rather high superconducting transition temperature ($T$$_{C}$ = 39 K) compared with other conventional superconductors \cite{J.Nagamatsu,X.Xi}. The first-principles calculations and the inelastic neutron scattering measurements revealed that the $E_{2g}$ in-plane boron phonons near the Brillouin zone center strongly coupled to the planar boron $\sigma$ bands \cite{K.Bohnen,T.Yildirim}, which leads to the high $T$$_{C}$ in MgB$_{2}$. Moreover, MgB$_{2}$ has the multiple bands with a weak electron correlation \cite{J.An,J.Kortus,I.Mazin}, and the distinct multiple superconducting energy gaps \cite{H.Choi,M.Iavarone}, resulting in markedly novel behaviors in its superconducting and normal-state properties \cite{X.Xi}. Although MgB$_{2}$ has been extensively studied, new physical phenomena are constantly discovered \cite{K.Jin,X.Zhou}, so those superconductors with the same structure are worth revisiting.

The binary bismuthide TlBi$_2$ \cite{E.Makarkov} crystalizes in a hexagonal AlB$_{2}$-type structure as the same as MgB$_{2}$, consisting of honeycombed bismuth layers and thallium layers located in between them, as shown in the inset of Fig. 1. Compared with MgB$_{2}$, TlBi$_{2}$ contains the heavier elements, suppressing the high frequency lattice vibration, and being unfavorable to the high $T$$_{C}$ superconductivity in the conventional electron-phonon coupling mechanism. Although TlBi$_{2}$ was classified into a strong coupling superconductor with $T$$_{C}$ = 6.4 K by R. Dynes in 1972 \cite{R.Dynes}, its detailed physical properties are rarely investigated as comparison with MgB$_{2}$.

In this article, we synthesized successfully the single phase polycrystalline TlBi$_2$ sample. The superconducting and normal state properties were systematically studied by resistivity, magnetization, Hall resistivity and specific heat measurements. We reconfirmed that type-II superconductivity with $T$$_{C}$ = 6.2 K, the upper critical field $\mu_{0}H_{c2}$ = 1.4 T, and the lower critical field $\mu_{0}H_{c1}$ = 1.08$\times$10$^{-2}$ T emerge in TlBi$_2$ compound. It was found that the electronic specific heat in the superconducting state can be well described using a single gap model within Bardeen-Cooper-Schrieffer (BCS) framework. The strong electron-phonon coupling occurs in this compound, confirmed by both the large $\lambda$$_{ep}$ (= 1.38) and the large $\Delta_{0}$/$k_{B}T_{C}$ (= 2.25) values, indicating that TlBi$_2$ is a conventional superconductor. It was found that the temperature dependence of resistivity in the normal state, $\rho$($T$), exhibits an unusual linear behavior above 50 K, which is ascribed to the low-energy phonon scattering, while $\rho$($T$) below 20 K is well described by the Fermi liquid theory, $i.e.$, $\rho(T) = \rho_{0} + AT^{2}$. Combining the specific heat data at normal state, we found that its Kadowaki-Woods ratio (KWR), $A/\gamma^{2}$ [9.2$\times$10$^{-5}$ $\mu\Omega$ cm(mol K$^{2}$/mJ)$^{2}$], is unexpectedly one order of magnitude larger than that obtained in many heavy Fermi compounds, although the electron correlation is not so strong.

\section{EXPERIMENTAL METHODS}

TlBi$_{2}$ polycrystalline samples were synthesized using the method as described before \cite{T.Claeson}. First, Tl and Bi lumps were mixed and sealed in a vacuum quartz tube. Then, the mixture was molten over a flame and mixed carefully by shaking vigorously for 10 min. After that, TlBi$_{2}$ samples were annealed at 210 $^{\circ}$C for 2 weeks. At last, the quartz tube was quenched into cold water to prevent the formation of impurity phases during cooling process. In order to compensate for the loss of Tl due to the presence of Tl$_{2}$O$_{3}$ in the raw material, an additional 10$\%$ Tl was added. The obtained sample was easy to press into a flake and then cut into rectangular bars for later study. Polycrystalline $x$-ray diffraction (XRD) was performed on a Rigaku $x$-ray diffractometer with Cu $K$$_{\alpha}$ radiation. The resistivity and Hall coefficient were measured using the standard four-probe technique. The heat capacity was measured using the relaxation method. All the transport properties were measured in a $Quantum Design$ Physical Properties Measurement System PPMS-9. The dc magnetization was obtained using a Magnetic Property Measurement System (\textit{Quantum Design} MPMS-VSM).

\section{RESULTS AND DISCUSSIONS}
\begin{figure}
  % Requires \usepackage{graphicx}
  \includegraphics[width=8cm]{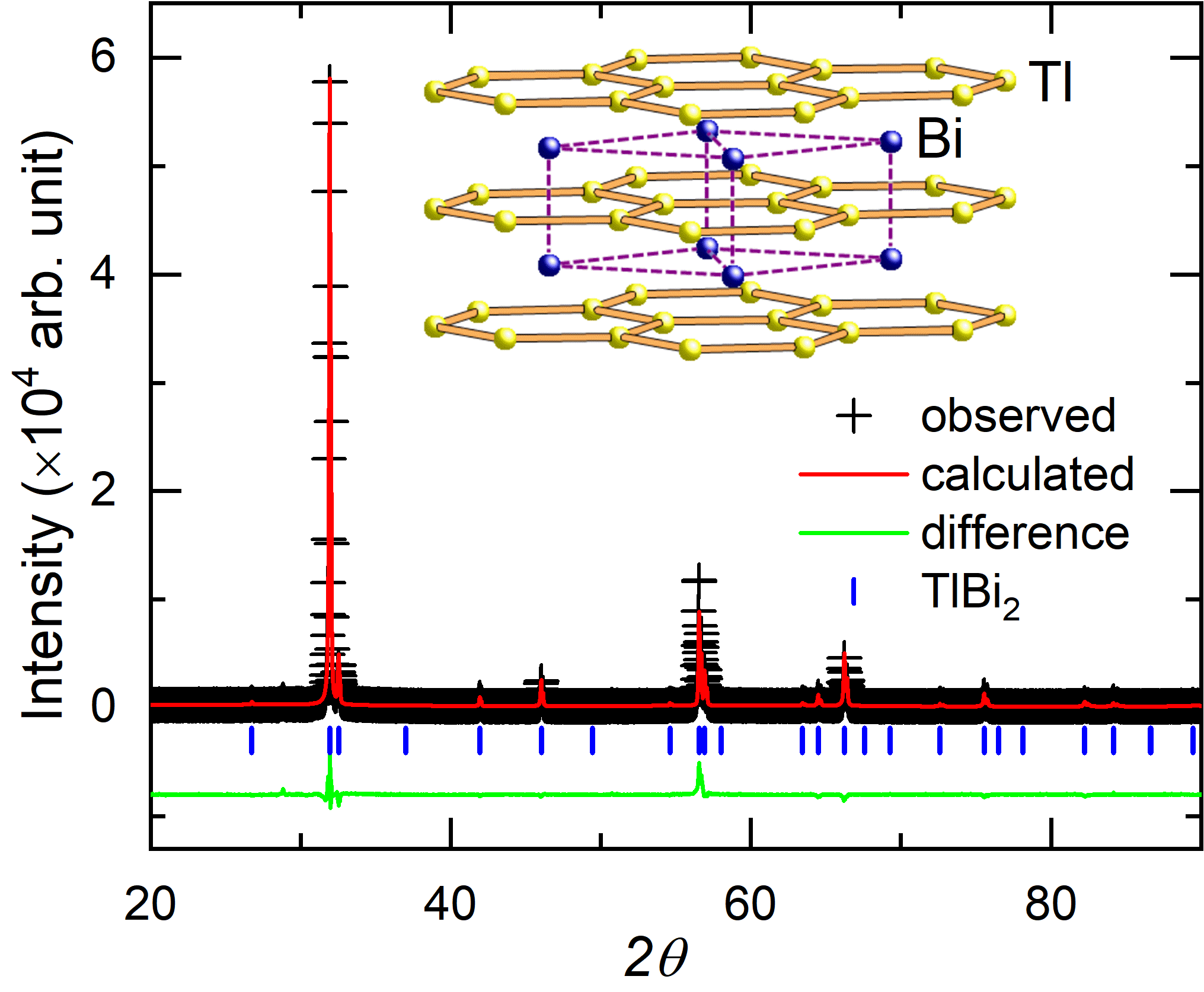}\\
  \caption{(Color online) Rietveld refinement profile of the polycrystalline XRD of TlBi$_{2}$. Inset: The crystal structure of TlBi$_{2}$. Thallium and bismuth atoms are drawn as dark blue and yellow spheres, respectively.}
\end{figure}

Figure 1 shows the XRD pattern of TlBi$_{2}$ sample. All the peaks can be well indexed with an AlB$_{2}$-type structure (space group: $P6/mmm$, $No.$ 191), and no obvious impurity peaks were observed. The lattice parameters $a$ = 5.68(2) \AA, and $c$ = 3.37(1) \AA, were obtained by Rietveld refinement, which is consistent with the results reported previously \cite{E.Makarkov}. The cell parameter $a$ of TlBi$_{2}$ is much larger than that of MgB$_{2}$ and $c$ is smaller, which originates from the nearly same Bi-Bi bond length along the $a$ and $c$ axis, implying its three-dimensional feature.

\begin{figure*}
 \includegraphics[width=14cm]{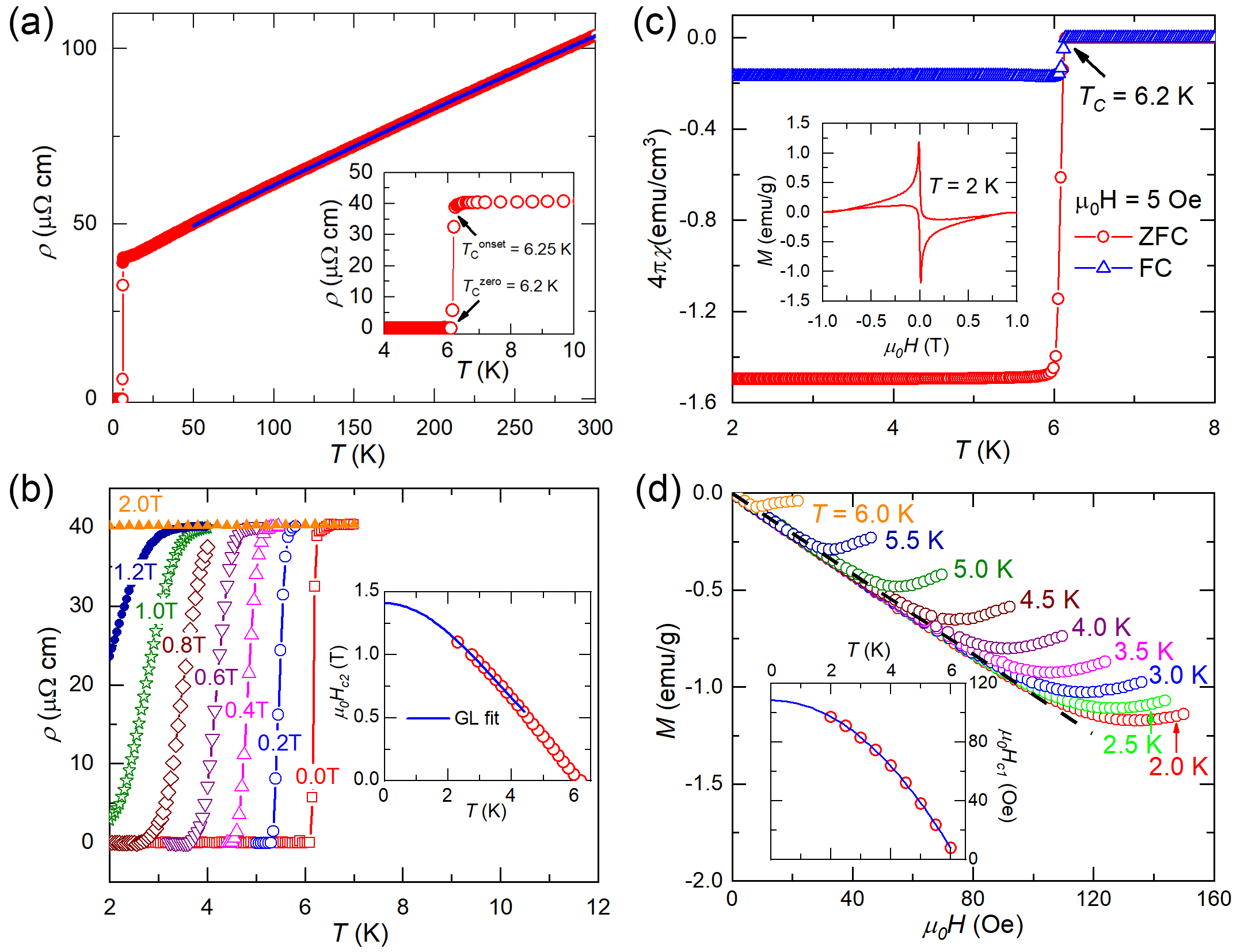}\\
 \caption{(color online) (a) Temperature dependence of resistivity of TlBi$_{2}$ between 2 and 300 K measured at zero field; Inset: The enlarge view near $T_{C}$; (b) Temperature dependence of resistivity under several selected magnetic fields below 8 K; Inset: The upper critical field $H$$_{c2}$ as a function of temperature for TlBi$_{2}$; (c) Temperature dependence of magnetic susceptibility below 8 K, measured at 5 Oe with both ZFC and FC processes; Inset: Field dependent magnetization $M$ measured between -1 and 1 T at 2 K; (d) Field dependent magnetization $M$ measured at various temperatures below 150 Oe. The dashed line indicates the initial linear magnetization curve. Inset: The lower critical field $H$$_{c1}$ as a function of temperature for TlBi$_{2}$.}
\end{figure*}

Figure 2(a) displays the temperature dependence of the resistivity, $\rho$($T$), between 2 and 300 K of TlBi$_{2}$ sample. The room temperature resistivity $\rho$(300K) is about 103.7 $\mu\Omega$ cm, close to that of TlSb \cite{Y.Zhou}. It's clear that TlBi$_{2}$ exhibits a metallic behavior in the whole measuring temperature range, $i.e.$, the resistivity decreases with decreasing temperature. At $T$$_{C}$$\mathrm{^{onset}}$ = 6.25 K, the resistivity drops abruptly to zero, suggesting that a bulk superconducting transition occurs with a transition width $\Delta$$T$$_{C}$ = 0.05 K, which is also confirmed by a large diamagnetic signal and a significant specific heat jump at $T$$_{C}$, as shown in Fig. 2(c) and Fig. 3, respectively. The superconducting transition temperature here is similar to that reported previously ($T$$_{C}$ = 6.4 K) \cite{R.Dynes}.

It is obvious that the $\rho$($T$) in the normal state exhibits a linear temperature dependence in a large region (50 K $\leq$ T $\leq$ 300 K), which can be ascribed to the low-energy phonon scattering here \cite{T.Takayama}, although the similar behavior in the cuprate high temperature superconductors was explained as the strong electron correlation effect. As shown by a blue line in the Fig. 2(a), we fitted the $\rho$($T$) data above 50 K using the standard Bloch-Gr$\ddot{u}$neisen (BG) formula,
\begin{equation}
  \rho = \rho_{0} + 4C(\frac{T}{\Theta_{D}})^{5} \int_{0}^{T/\Theta_{D}}\frac{x^{5}}{(e^{x}-1)(1-e^{-x})} \textrm{d} x
\end{equation}
then we obtained the residual resistivity $\rho_{0}$ = 41 $\mu\Omega$ cm, the fitting parameter $C$ = 17.5 $\mu\Omega$ cm/K, and the Debye temperature $\Theta_{D}$ = 83 K. We also found that the $\rho$($T$) below 50 K deviates from the $T$-linear dependence and turns to $T$-square dependence below 20 K, indicating the Fermi-liquid ground state, discussed in details as follows.

In order to obtain the upper critical field $H$$_{c2}$($T$), we measured the resistivity at various magnetic fields between 2 K and 8 K, as shown in Fig. 2(b). With increasing magnetic field, the superconducting transition shifts to lower temperature. At 2.0 T, the superconducting transition is not observed above 2 K. The $H$$_{c2}$ is determined by the temperature when the resistivity drops to 50\% of the normal state value and is plotted as a function of temperature in the inset of Fig. 2(b). According to the Ginzburg-Landau (GL) theory, the $\mu_{0}$$H$$_{c2}$ value at zero temperature was estimated to be 1.4 T using the formula
\begin{equation}
 H_{c2}(T) = H_{c2}(0)(1-t^{2})/(1+t^{2})
\end{equation}
where $t$ is the reduced temperature $T/T_{C}$. Then the coherence length $\xi$$_{GL}$ of TlBi$_{2}$ was estimated to be 15.3 nm from the relation, $\xi_{GL}^{2} = \Phi_{0}/2\pi H_{c2}(0)$, where $\Phi$$_{0}$ = $h/2e$ is the magnetic flux quantum ($\approx$ 2.07$\times$10$^{-15}$ Wb).

Figure 2(c) shows the temperature dependence of the magnetic susceptibility, $\chi$($T$), measured at an applied field of 5 Oe both in the zero field cooling (ZFC) and field-cooling (FC) processes. A sharp superconducting transition and a quite flat feature below $T$$_{C}$ are clearly observed in $\chi$($T$), suggesting superconductivity emerges in the sample. At $T$ = 2 K, the 4$\pi\chi$ value exceeds -100\% eum/cm$^{3}$ due to the demagnetization effect. The $M$($H$) curve measured at $T$ = 2 K, as shown in the inset of Fig. 2(c), exhibits a typical type-II superconducting behavior.

To obtain the lower critical field $H$$_{c1}$($T$), we measured the $M$($H$) curves at different temperatures, as shown in Fig. 2(d). The $H$$_{c1}$($T$) determined by the field where $M$ starts to deviate from the initial linear curve, is plotted as a function of temperature in the inset of Fig. 2(d). It is clear that the $H$$_{c1}$ can be well described using the GL theory as

\begin{equation}
  H_{c1}(T) = H_{c1}(0)[1-(\frac{T}{T_{C}})^2]
\end{equation}

The lower critical field at zero temperature, $H_{c1}$(0), was estimated to be 108 Oe. The penetration depth $\lambda$$_{GL}$ was estimated to be 198 nm using the relation $H_{c1}(0) = \frac{\Phi_{0}}{4\pi\lambda^{2}_{GL}}ln(\frac{\lambda_{GL}}{\xi_{GL}})$ and the $\xi_{GL}$ value obtained above. Then, the GL parameter, $\kappa$$_{GL}$ = $\lambda_{GL}/\xi_{GL}$, was calculated to be 12.9, much larger than 1/$\sqrt{2}$, confirming that TlBi$_{2}$ is a type-II superconductor. The thermodynamic critical field $H_{c}$(0) was also estimated to be 770 Oe from the relation, $H_{c1}(0)H_{c2}(0) = H_{c}^2(0)ln\kappa_{GL}$, which is almost an order of magnitude smaller than that of MgB$_{2}$ \cite{X.Xi}.

\begin{figure}
  % Requires \usepackage{graphicx}
  \includegraphics[width=8cm]{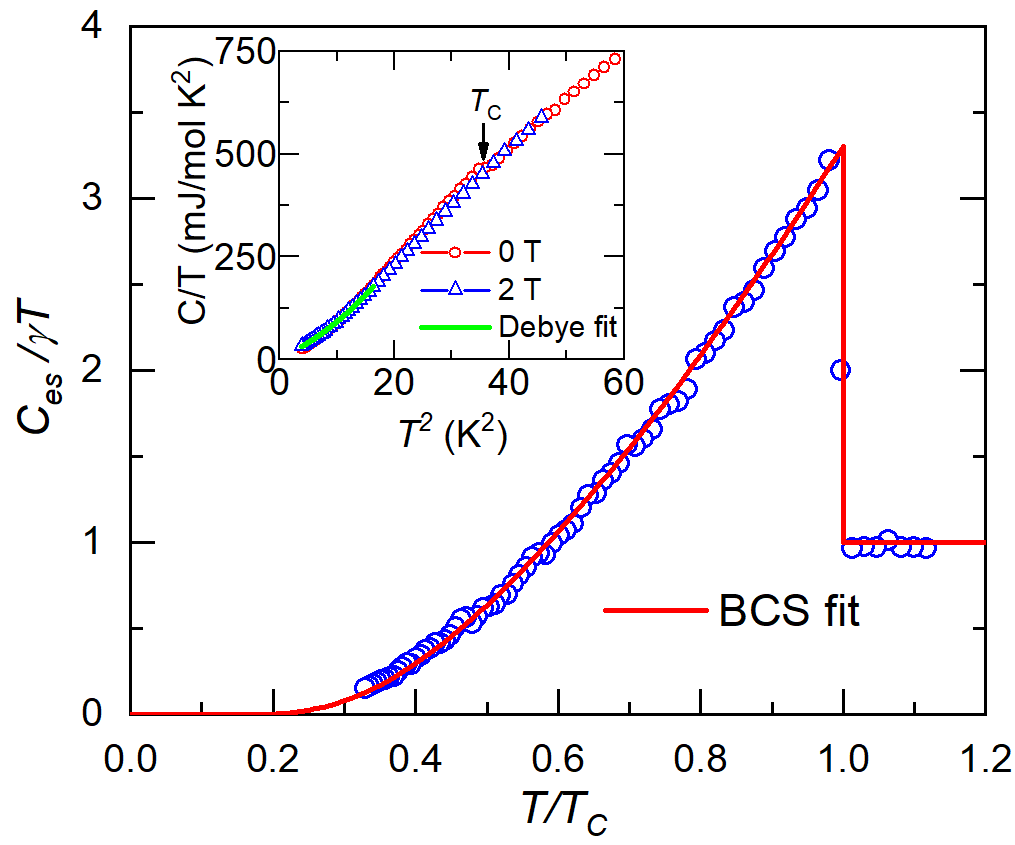}\\
  \caption{(Color online) The electronic specific heat divided by the product of Sommerfeld coefficient $\gamma$ and temperature as a function of the reduced temperature $T/T_{C}$ in the superconducting state at zero field; Inset: temperature square dependence of $C/T$, measured at 0 T and 2 T magnetic fields. The green line is the fit to the data as described in the text.}
\end{figure}

To get the information of superconducting transition, we also carried out the specific heat, $C$($T$), measurements at both 0 T and 2 T. The inset of Fig. 3 shows the temperature square dependence of $C/T$ with a small difference between the two curves below $T_{C}$. The low temperature $C$($T$) measured at 2 T, where bulk superconductivity is completely suppressed, can be well fitted using the Debye model, $C/T = \gamma + \beta T^2 + \delta T^4$, where $\gamma$ is the Sommerfeld coefficient, $\beta$ Debye constant, and $\delta$ the fitting parameter. The first and last two terms are ascribed to the electronic and phonon contribution, respectively. We obtained $\gamma$ = 8.63 mJ/(mol K$^{2}$), $\beta$ = 4.87 mJ/(mol K$^{4}$), and $\delta$ = 0.35 mJ/(mol K$^{6}$) by the best fit to the data below 4 K (the solid green line). Then, the Debye temperature $\Theta_{D}$, was evaluated to be 104 K, which is close to that (83 K) obtained from the $\rho$($T$) data mentioned above, from the relation $\Theta_{D} = (12\pi^{4}RN/5\beta)^{1/3}$, where $R$ = 8.31 J/(mol K) is the molar gas constant, and $N$ is the number of atoms per unit cell. The electronic specific heat, $C_{es}$($T$), in the superconducting state was obtained by subtracting the phonon contribution from the total $C$($T$). Figure 3 presents the $C_{es}$/$\gamma$$T$ $vs$. $T/T_{C}$, a sharp jump of 2.32 emerging at $T_{C}$, which is significantly larger than that of the well known BCS theory (1.43), suggesting that the strong electron-phonon coupling occurs in TlBi$_{2}$. The electron-phonon coupling constant, $\lambda$$_{ep}$, can be derived from the modified McMillan formula \cite{P.Allen,J.Carbotte,T.Klimczuk}

\begin{equation}
 \lambda_{ep} = \frac{1.04 + \mu^{*}ln(\frac{\omega_{ln}}{1.2T_{C}})}{(1-0.62\mu^{*})ln(\frac{\omega_{ln}}{1.2T_{C}})-1.04}
\end{equation}

where $\mu^{*}$ is the Coulomb pseudopotential, which has been reported to be 0.121 \cite{R.Dynes}, and $\omega_{ln}$ is the logarithmic averaged phonon frequency, which can be estimated from the specific heat jump at $T_{C}$ using the formula \cite{P.Allen,J.Carbotte,T.Klimczuk}

\begin{equation}
 \frac{\Delta C}{\gamma T_{C}} = 1.43\times[1+53(\frac{T_{C}}{\omega_{ln}})^{2}ln(\frac{\omega_{ln}}{3T_{C}})]
\end{equation}

Taking $\Delta C/\gamma T_{C}$ = 2.32 and $T_{C}$ = 6.2 K, we obtained $\omega_{ln}$ = 63.3 K and $\lambda_{ep}$ = 1.38, which is smaller than $\lambda_{ep}$ = 1.63 reported previously \cite{R.Dynes}. However, it is still large compared with those for typical strong coupling superconductors such as Mo$_{6}$Se$_{8}$ ($\lambda_{ep}$ = 1.27), and Pb-Tl alloy ($\lambda_{ep}$ = 1.15-1.53) \cite{P.Allen}, indicating the strong-coupling nature of superconducting pairing.

Then, we analysed the electronic specific heat data $C_{es}$($T$) using BCS theory with a single gap. Within the framework of BCS theory, the thermodynamic properties, entropy ($S$) and electronic specific heat ($C_{es}$), can be written as

\begin{equation}\label{}
S =-\frac{6\gamma}{\pi^{2}} \frac{\Delta_{0}}{k_{B}}\int_{0}^{\infty}[f\ln{f}+(1-f)\ln{(1-f)}] \textrm{d} y
\end{equation}
\begin{equation}\label{}
C_{es}=T\frac{\textrm{d}S}{\textrm{d}T}
\end{equation}

\begin{table}
\caption{\label{Table1}%
The superconducting parameters for TlBi$_{2}$ superconductor.
}
\begin{ruledtabular}
\begin{tabular}{cccc}
\textrm{} &
\textrm{Parameters (unit)}&
\textrm{Value}&
\textrm{}\\
\colrule
& $T_{C}$ (K) & 6.2 & \\
& $\mu$$_{0}$$H_{c1}$(0) (T) & 1.08$\times$10$^{-2}$ & \\
& $\mu$$_{0}$$H_{c2}$(0) (T) & 1.4 & \\
& $\xi$$_{GL}$ (nm) & 15.3 & \\
& $\lambda$$_{GL}$ (nm) & 198 & \\
& $\kappa$$_{GL}$ & 12.9 & \\
& $\lambda$$_{ep}$ & 1.38 & \\
& $\gamma$ (mJ/mol K$^{2}$) & 8.63 & \\
& $\Delta C_{es}/\gamma T_{C}$ & 2.32 & \\
& $\Delta_{0}/k_{B}T_{C}$ & 2.25 & \\
\end{tabular}
\end{ruledtabular}
\end{table}

where $f=[\exp(\beta E)+1]^{-1}$ and $\beta = (k_{B}T)^{-1}$, whereas the integration variable is $y = \varepsilon/\Delta_{0}$. The energy of the quasiparticles is evaluated from the relation $E = [\varepsilon^{2}+\Delta^{2}_{0}\delta^{2}(t)]^{0.5}$, where $\varepsilon$ is electron energy with respect to the Fermi energy and $\delta(t)$ is the normalized BCS gap at the reduced temperature $t = T/T_{C}$ as tabulated by M$\ddot{u}$hlschlegel \cite{B.Muhlschlege}. As shown in the figure, the single-gap model presents a good fit to $C_{es}/\gamma T$, suggesting that TlBi$_{2}$ is a phonon-mediated $s$-wave superconductor. Meanwhile, the $\Delta$$_{0}$/$k_{B}T_{C}$ was fitted to be 2.25, agrees well with that obtained from the tunneling experiments\cite{P.Vashishta}, which is much larger than that as predicted for a weak coupling limit ($\Delta$$_{0}$/$k_{B}T_{C}$ = 1.76), further confirming the strong electron-phonon coupling in TlBi$_{2}$. The obtained superconducting parameters are summarized in Table I.

\begin{figure}
  % Requires \usepackage{graphicx}
  \includegraphics[width=8cm]{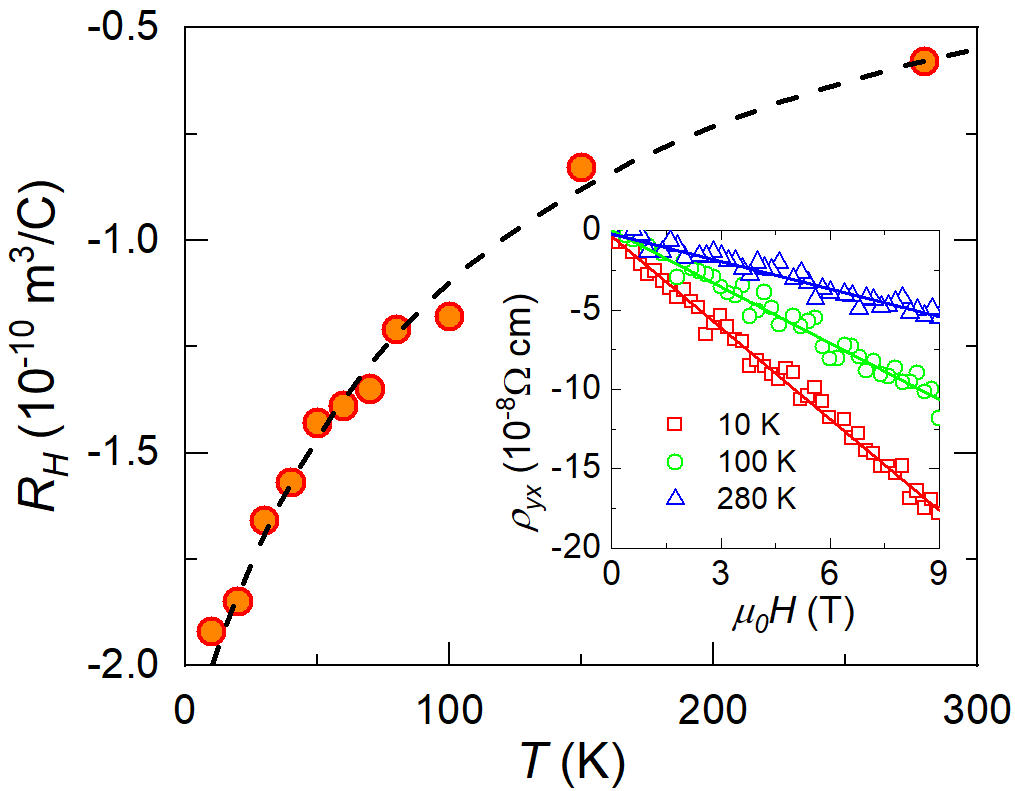}\\
  \caption{(Color online) Temperature dependence of the Hall coefficient $R$$_{H}$ of TlBi$_{2}$ between 10 and 280 K. The dashed line is a guide for eyes. Inset: Isothermal Hall resistivity at $T$ = 10, 100, and 280 K.}
\end{figure}

Figure 4 shows the temperature dependence of the Hall coefficient $R$$_{H}$ of TlBi$_{2}$. The transverse Hall resistivity, $\rho$$_{yx}$, was derived from the antisymmetric part of the transverse resistivity under the reversal of magnetic field at a given temperature. As shown in the inset of Fig. 4, the $\rho$$_{yx}$ exhibits a linear dependence with the magnetic field below 9 T, suggesting that $R$$_{H}$ is independent of the magnetic field. At $T$ = 10 K, $R$$_{H}$ is about -1.9$\times$10$^{10}$ m$^{3}$/C, indicating that the dominant carriers are electron-type. With increasing temperature, $R$$_{H}$ increases monotonically, but remains negative below 300 K. The significant $T$-dependent behavior of $R$$_{H}$ is similar to that observed in MgB$_{2}$ \cite{R.Jin}. If we assume that the Drude relation holds for TlBi$_{2}$ even in the case of multiple bands, the carrier concentration $n$ could be estimated to be 3.3$\times$10$^{22}$/cm$^{3}$ at $T$ = 10 K. Assuming a parabolic dispersion with spherical Fermi surface, the Fermi wave vector, $k_{F}$, could be calculated to be 9.9$\times$10$^{9} $m$^{-1}$ from $k_{F} = (3n\pi^{2})^{1/3}$. Then the band Sommerfeld coefficient $\gamma_{b}$ was calculated to be 3.0 mJ/(mol K$^{2}$) from the relation $\gamma_{b} = \pi^{2}nk_{B}^{2}m_{e}/\hbar^{2}k_{F}^{2}$. In a strong coupling compound, the electronic specific heat coefficient $\gamma$$_{cal}$ is expected to enhance to be (1+$\lambda_{ep}$)$\gamma_{b}$ due to electron-phonon coupling \cite{T.Takayama,M.Bruhwiler}, where $\lambda_{ep}$ is the electron-phonon coupling constant. Using the value for $\lambda_{ep}$ (1.38) determined from our measurements, the $\gamma$$_{cal}$ is calculated to be 7.14 mJ/(mol K$^{2}$), which is close to that obtained from the specific heat measurements, indicating that the Fermi-liquid theory can well described the behavior of TlBi$_{2}$ compound.

\begin{figure}
  % Requires \usepackage{graphicx}
  \includegraphics[width=8cm]{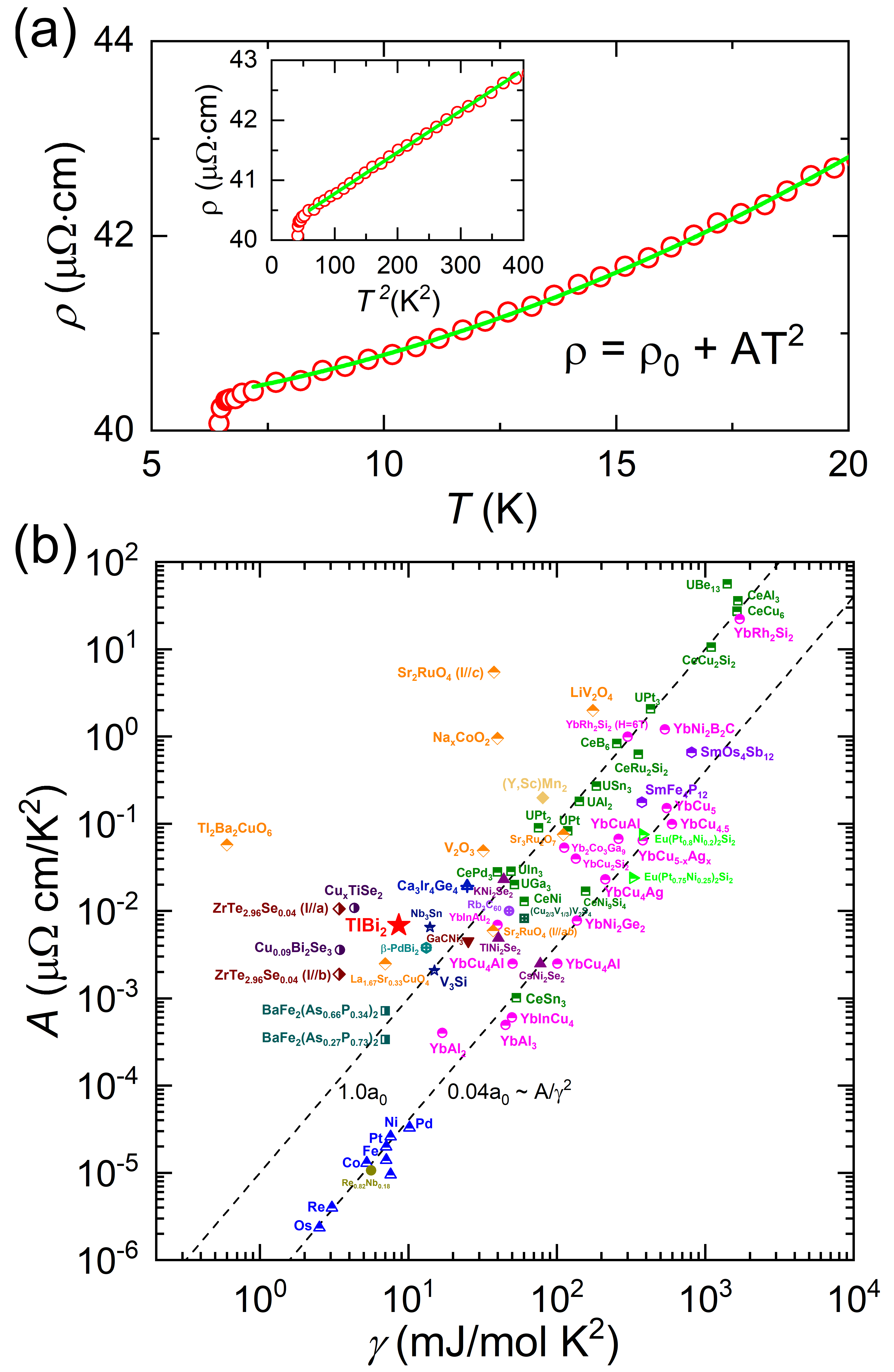}\\
  \caption{(Color online) (a) Temperature dependence of resistivity below 20 K; Inset: temperature square dependence of resistivity below 20 K; The solid green lines correspond to $\rho = \rho_{0} +AT^{2}$, just as mentioned in the main text; (b) The coefficient $A$ $vs$. the Sommerfield coefficient $\gamma$ for various compounds. The data beyond TlBi$_{2}$ were collected from previous papers \cite{Y.Fang}, including CDW materials \cite{K.Wagner,X.Zhu}, oxides \cite{Y.Maeno,S.Li,C.Urano,S.Nakamae,A.Jacko,N.Tsujii}, A-15 superconductors \cite{K.Miyake}, heavy fermions \cite{K.Kadowaki,N.Tsujii,N.Tsujii2}, transition metals \cite{M.Rice}, and so on \cite{H.Wang,H.Chen}.}
\end{figure}

The KWR compares the temperature dependence of a metal's resistivity to that of its specific heat \cite{K.Kadowaki,M.Rice}, thereby probing the relationship between the electron-electron scattering rate and the renormalization of the electron mass, which is considered as a measurement of the electron-electron correlation strength. To deduce the KWR value of TlBi$_{2}$, we fitted the $\rho$($T$) data between 7 and 20 K using the Fermi-liquid prediction, $\rho = \rho_{0} + AT^{2}$, when electron-electron scattering dominates over electron-phonon scattering. The residual resistivity $\rho_{0}$ = 40.1 $\mu\Omega$ cm and the coefficient $A$ = 6.84$\times$10$^{-3}$ $\mu\Omega$ cm/K$^{2}$ were obtained by the fitting, as shown in Fig. 5(a). Using the Sommerfeld coefficient $\gamma$ = 8.63 mJ/(mol K$^{2}$) obtained from $C$($T$) measurement, we obtained the KWR of TlBi$_{2}$ to be 9.2$\times$10$^{-5}$ $\mu\Omega$ cm(mol K$^{2}$/mJ)$^{2}$. For many heavy-fermion compounds, the KWR value is close to 1.0$\times$10$^{-5}$ $\mu\Omega$ cm(mol K$^{2}$/mJ)$^{2}$ = 1.0$a$$_{0}$, while for a lot of transition metals, the KWR value is close to 0.04$\times$10$^{-5}$ $\mu\Omega$ cm(mol K$^{2}$/mJ)$^{2}$ = 0.04$a$$_{0}$ \cite{K.Kadowaki,A.Jacko}, where $a$$_{0}$ = 10$^{-5}$ $\mu\Omega$ cm(mol K$^{2}$/mJ)$^{2}$. For a comparison, we plot the KWR for various compounds in Fig. 5(b). It is obvious that the KWR value of TlBi$_{2}$ obtained here is somehow unexpectedly larger than that obtained in many heavy Fermi compounds, because the electron correlation seems not so strong as discussed above. It has been argued that the KWR in the heavy fermions is larger than that in transition metals due to the stronger correlation \cite{K.Miyake}. However, several scenarios have been proposed to explain the large KWRs observed in UBe$_{13}$, transition metals oxides and organic charge-transfer salts, including scattering \cite{K.Miyake}, proximity to a quantum critical point \cite{S.Li}, as well as the suggestion that electron-phonon scattering in reduced dimensions may result in a quadratic temperature dependence of the resistivity \cite{C.Strack}. Especially, Matsuura $et$ $al.$ \cite{T.Matsuura,C.Yu} suggested that the strong dynamical coupling between conduction electrons and phonons may give rise to the heavy fermi bands at low temperatures. Thus, the strong coupling compounds may obey the universal KWR of heavy fermi compounds, $i.e.$, $A$/$\gamma$$^{2}$ $\approx$ 10$^{-5}$ $\mu\Omega$ cm(mol K$^{2}$/mJ)$^{2}$. Although we can not determine which mechanism can account for the unexpectedly large KWR in TlBi$_{2}$, it is worth noting that our KWR value is almost an order of magnitude larger than that obtained in many heavy Fermi compounds, suggesting that this compound may provide a novel material platform for the study of large KWR.

\section{CONCLUSION}

In summary, we systematically investigated the superconducting and normal state properties of TlBi$_{2}$. Type-II superconductivity with $T_{C}$ = 6.2 K, the upper critical field $\mu_{0}H_{c2}$ = 1.4 T, and the lower critical field $\mu_{0}H_{c1}$ = 1.08$\times$10$^{-2}$ T was revealed by the resistivity and magnetization measurements. It was found that the electronic specific heat in the superconducting state can be well described using a single gap model within the BCS framework. The strong electron-phonon coupling occurs in this compound, confirmed by both the large $\lambda$$_{ep}$ (1.38) and the large $\Delta_{0}$/$k_{B}T_{C}$ (2.25) values, indicating that TlBi$_2$ is a conventional superconductor. In the normal state, the resistivity exhibits $T$-linear dependence above 50 K, which is ascribed to the low-energy phonon scattering, and $T$-square dependence below 20 K, suggesting the Fermi-liquid ground state. Combining the specific heat data at normal state, a large KWR [9.2$\times$10$^{-5}$ $\mu\Omega$ cm(mol K$^{2}$/mJ)$^{2}$] was obtained, which is an order of magnitude larger than that obtained in many heavy Fermi compounds, although the electron correlation is not so strong.

\begin{acknowledgments}
This work was supported by the Ministry of Science and Technology of China under Grant No. 2016YFA0300402 and the National Natural Science Foundation of China (NSFC) (Nos. 11974095, 12074335), and the Fundamental Research Funds for the Central Universities.

\end{acknowledgments}

\bibliography{TlBi2}

%\begin{thebibliography}{unsrt}

%\end{thebibliography}

\end{document}